\documentclass[conference, 10pt]{IEEEtran}

\usepackage{cite}
\usepackage{url}
\usepackage{graphicx}
\usepackage{color}
\usepackage{placeins}
\usepackage{float}
\usepackage{tabularx,colortbl}
\usepackage{ifthen}
\usepackage[font={small}]{caption}
\usepackage[font={small}]{subcaption}
\usepackage{tikz}

\makeatletter

\newcounter{author}
\renewcommand{\author}[2][]{
   \stepcounter{author}
   \@namedef{author@\theauthor}{#2}
   \@namedef{authorlabel@\theauthor}{#1}
}

\newcounter{address}
\newcommand{\address}[2][]{
   \stepcounter{address}
   \@namedef{address@\theaddress}{#2}
   \@namedef{addresslabel@\theaddress}{#1}
}

\newcommand{\alsep}{and}

\def\newmaketitle{\par%
  \begingroup%
  \normalfont%
  \def\thefootnote{}
  \def\footnotemark{}
  \let\@makefnmark\relax
  \footnotesize
  \footnotesep 0.7\baselineskip
  \normalsize%
  \twocolumn[\thenewmaketitle\@IEEEaftertitletext]%
  \if@IEEEusingpubid
     \enlargethispage{-\@IEEEpubidpullup}%
  \fi
  \endgroup
  \setcounter{footnote}{0}\let\maketitle\relax\let\@maketitle\relax
  \gdef\@thanks{}%
  \let\thanks\relax}

\def\thenewmaketitle{
  \newpage
  \begin{center}%
    \vskip0.2em{\Huge\@IEEEcompsoconly{\sffamily}\@IEEEcompsocconfonly{\normalfont\normalsize\vskip 2\@IEEEnormalsizeunitybaselineskip
   \bfseries\large}\@title\par}\vskip1.0em\par%
    \vspace{1ex}
    \newcounter{c@author}
    \newcounter{c@tmp}
    \ifthenelse{\value{author}=2}{%
      \newcommand{\liand}{ and }}{%
      \newcommand{\liand}{, and }}
    \ifthenelse{\value{address}<2}{%
      \@nameuse{author@1}%
      \stepcounter{c@author}%
      \whiledo{\value{c@author}<\value{author}}{%
        \setcounter{c@tmp}{\value{author}}%
        \addtocounter{c@tmp}{-\value{c@author}}%
        \ifthenelse{\value{c@tmp}=1}{%
          \renewcommand{\alsep}{\liand}}{\renewcommand{\alsep}{, }}%
        \stepcounter{c@author}\alsep \@nameuse{author@\thec@author}}\\%
    }
    {
      \@nameuse{author@1}${}^{(\ref{\@nameuse{authorlabel@1}})}$%
      \stepcounter{c@author}%
      \whiledo{\value{c@author}<\value{author}}{%
      \setcounter{c@tmp}{\value{author}}%
      \addtocounter{c@tmp}{-\value{c@author}}%
      \ifthenelse{\value{c@tmp}=1}{%
        \renewcommand{\alsep}{\liand}}{\renewcommand{\alsep}{, }}%
      \stepcounter{c@author}\alsep \@nameuse{author@\thec@author}%
        ${}^{(\ref{\@nameuse{authorlabel@\thec@author}})}$%
      }
    }
    \vspace{0.2ex}

    \ifthenelse{\value{address}>0}{%
      \ifthenelse{\value{address}=1}{
        {\@nameuse{address@1}}
      }
      {
        \newcounter{c@address}

        \begin{center}
        \whiledo{\value{c@address}<\value{address}}
        {
          \refstepcounter{c@address}
            ${}^{(\thec@address)}$\,%
              \label{\@nameuse{addresslabel@\thec@address}}%
              \@nameuse{address@\thec@address}\\ %
        }
        \end{center}
      } 
    }
    {
      \relax
    }
  \end{center}
}

\makeatother

\title{Computationally-tractable synthesis of an MXene metamaterial absorber with a 3D-printable spatially variable substrate by a local approach }

\author[org1]{Maria-Thaleia Passia}
\author[org1]{Yilin Zhao}
\author[org1]{Haozhe Wang}
\author[org1]{Steven Cummer}

\address[org1]{Department of Electrical and Computer Engineering, Duke University, NC, USA}

\newsavebox{\fundinglogo}
\sbox{\fundinglogo}{%
    \begin{tikzpicture}[baseline, overlay]
        \node[anchor=center] at (5em,-2pt) {\includegraphics[width = 0.4\columnwidth]{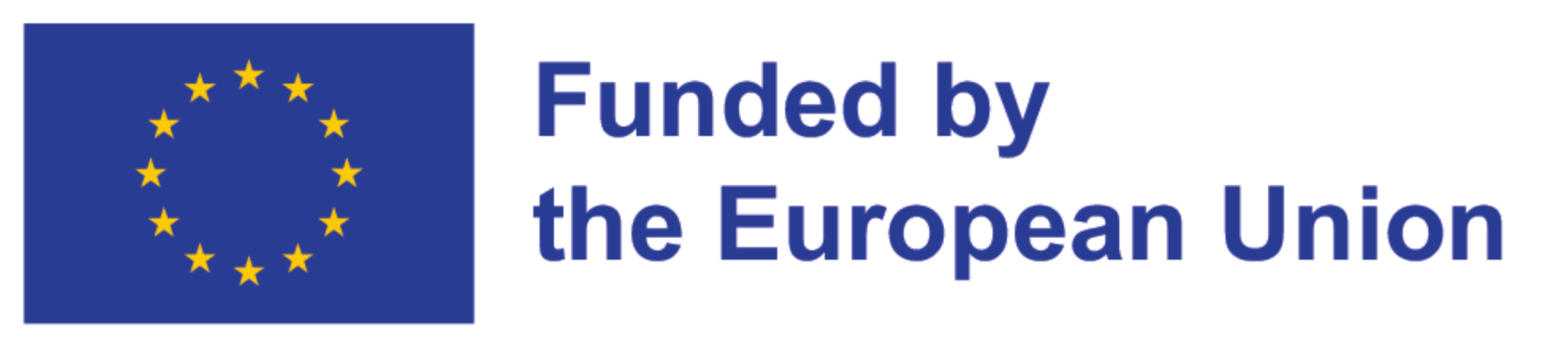}}; 
    \end{tikzpicture}%
}

\let\svthefootnote\thefootnote

\newcommand\blankfootnote[1]{%
  \let\thefootnote\relax\footnotetext{#1}%
  \let\thefootnote\svthefootnote%
}

\begin{document}

\newmaketitle

\begin{abstract}   
We introduce the LOCABINACONN3D methodology to enable the computationally tractable synthesis of MXene metamaterial absorbers (MMA) based on spatially variable 3D-printable substrates.  
Spatially variable substrates offer enhanced absorption bandwidth compared to constant ones. Such MXene MMAs are typically synthesized by inverse design, which may lead to non-manufacturable optimized dielectric substrates. To transform non-manufacturable dielectric substrates into manufacturable ones, existing methodologies either add manufacturing constraints to the optimization, which may lead to less optimized MMAs, or are computationally expensive, as they require full-wave simulations of the entire manufacturable
MMA.
We develop a computationally tractable methodology, LOCABINACONN3D, to render optimized MMAs manufacturable, preserving performance. 
Our methodology  (i) 
accommodates detailed connectivity constraints across consecutive layers, thus facilitating  multilayer fabrication, 
and (ii) enables scaling to larger MMAs, by  requiring simulations  only of smaller manufacturable MMA subareas
 and by using a semi-analytical method-of-lines (MoL) solver instead of full-wave methods to determine suitable manufacturable configurations.
This work paves the way for synthesizing optimized larger-scale 3D-printable MMAs more efficiently.
\end{abstract}

\section{Introduction}
Broadband absorption is essential in various applications, such as solar energy harvesting, radar cross-section (RCS) reduction, and electromagnetic interference shielding.
Metamaterial absorbers (MMA) can achieve broadband absorption by superimposing multiple resonant modes with different resonance frequencies~\cite{Fan2020}. Non-uniform multi-element MMAs can achieve broadband absorption~\cite{Liu2012}, with the resonators placed on the same plane or across different layers, forming a multilayer MMA. Multilayer MMAs are traditionally based on printed circuit boards (PCBs)~\cite{Ding2012}, which limits absorbance due to copper's relatively low ohmic losses, and has a complex stacking process. 
MMAs based on conductive materials with higher ohmic losses than copper  have recently been realized, such as conductive inks~\cite{Ghosh2018}, filaments~\cite{Lleshi2021}, indium tin oxide (ITO)~\cite{Chen2025}, MXenes~\cite{Feng2024,Zhang2025}. Among these alternatives, MXenes alleviate the need for additives and post-treatment in metal inks and are considerably thinner than conductive filaments, making them attractive candidates for MMA design.  

\blankfootnote{Copyright (C) 2026 IEEE. Personal use of this material is permitted.  Permission from IEEE must be obtained for all other uses, in any current or future media, including reprinting/republishing this material for advertising or promotional purposes, creating new collective works, for resale or redistribution to servers or lists, or reuse of any copyrighted component of this work in other works.}

To further enhance the absorbance bandwidth of multilayer MXene-based MMAs, spatially variable dielectric substrates of complex optimized shapes have recently been used instead of constant ones~\cite{Passia2025PRA}.  Tailoring the underlying physics of the dielectric structures on a deep subwavelength scale offers increased design flexibility and enhanced absorption. The spatially variable dielectric structures are fabricated by stereolithography (SLA) 3D-printing, which offers high precision and ease of fabrication,  and are made of resin and air. To synthesize such spatially variable complex structures, a density-based topology optimization is used,  the \texttt{TopOpt} method~\cite{Christiansen2021}.  The \texttt{TopOpt} method uses adjoint sensitivity analysis to perform gradient computations and results in structures, where the dielectric constant continuously varies within air and resin. Binarization and connectivity constraints can be added to the \texttt{TopOpt}; however, this may lead to less optimized devices compared to omitting these constraints. A simple post-processing methodology, the \texttt{BINACONN3D} methodology,  was introduced as an alternative~\cite{Passia2025PRA}. The \texttt{BINACONN3D} methodology transforms the optimized device of a continuous dielectric profile into a manufacturable one, while preserving the device’s performance~\cite{Passia2024}. \texttt{BINACONN3D} assigns a resin percentage to each material instead of a predetermined air/resin configuration, which provides greater design flexibility. It additionally incorporates fabrication constraints that facilitate the assembly of multilayered complex-shaped structures.  However, the selection of suitable air-resin structures is performed on a device level, requiring computationally-expensive full-wave simulations of the entire manufacturable MMA. As the device complexity increases, a need for more scalable approaches arises.

In this work, we introduce the \texttt{LOCABINACONN3D} methodology, which is a computationally tractable and scalable alternative to the 
\texttt{BINACONN3D}. The selection of suitable air-resin structures is performed on smaller MMA components instead of a device level; hence, we alleviate the need of computationally demanding simulations of the entire structure. 
Additionally, \texttt{LOCABINACONN3D} uses a semi-analytical method of lines (MoL) formulation instead of a full-wave method, which accelerates the simulation of the manufacturable components. As proof of concept, we synthesize the optimized manufacturable MMA by \texttt{LOCABINACONN3D} and compare its performance to that obtained by \texttt{BINACONN3D}. The advantages of the proposed method will be more pronounced in even larger-scale devices.

\begin{figure}
  \centering
    \subfloat[]{\label{fig:geom_constant}%
    \includegraphics[width = 0.8\columnwidth, trim = 0cm 1.2cm 0cm 0cm, clip]{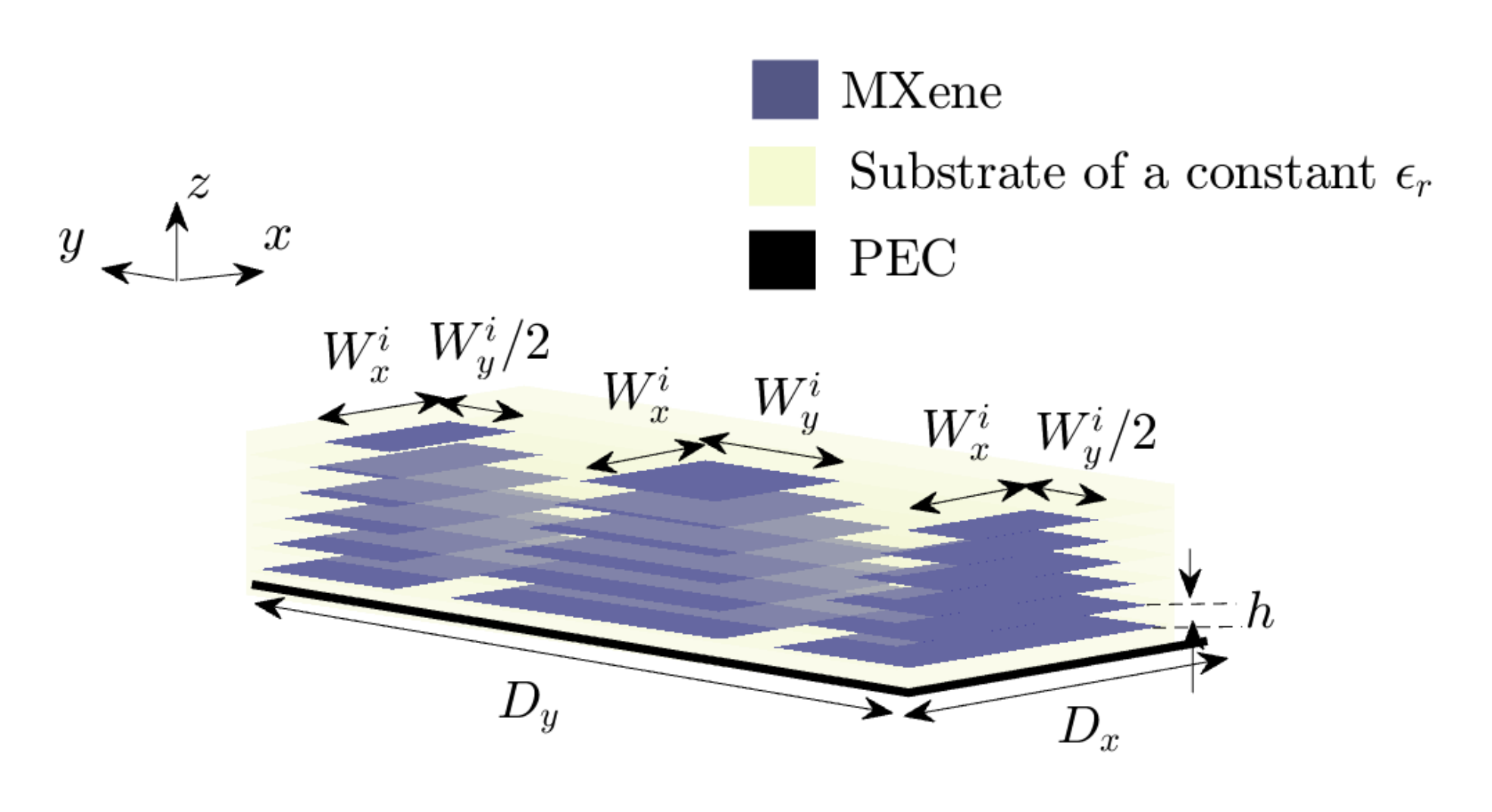}
    }
    \\
    \subfloat[]{\label{fig:opt}%
    \includegraphics[width = 0.9\columnwidth]{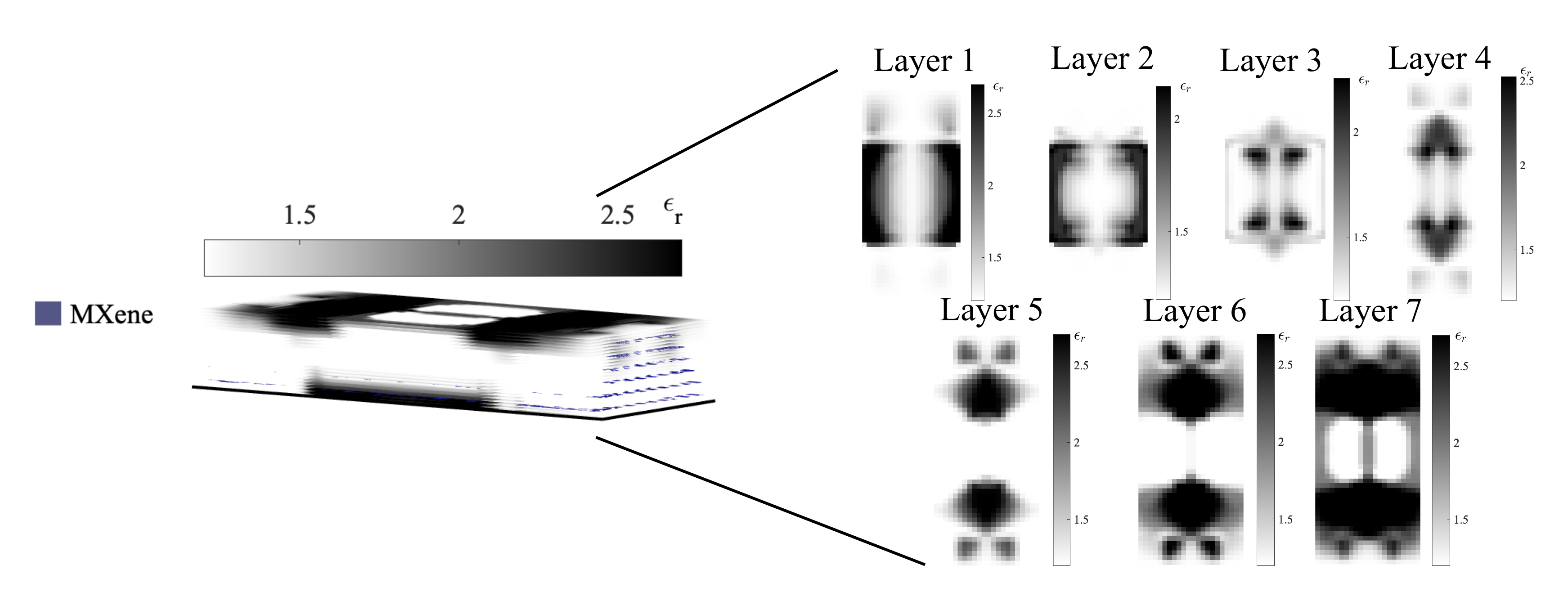}
      }   
    \caption{(a) MXene MMA of constant dielectric substrates. (b) Optimized MXene MMA with spatially variable dielectric substrates.}
\end{figure}

\begin{figure}
    \centering
    \includegraphics[width=1\linewidth]{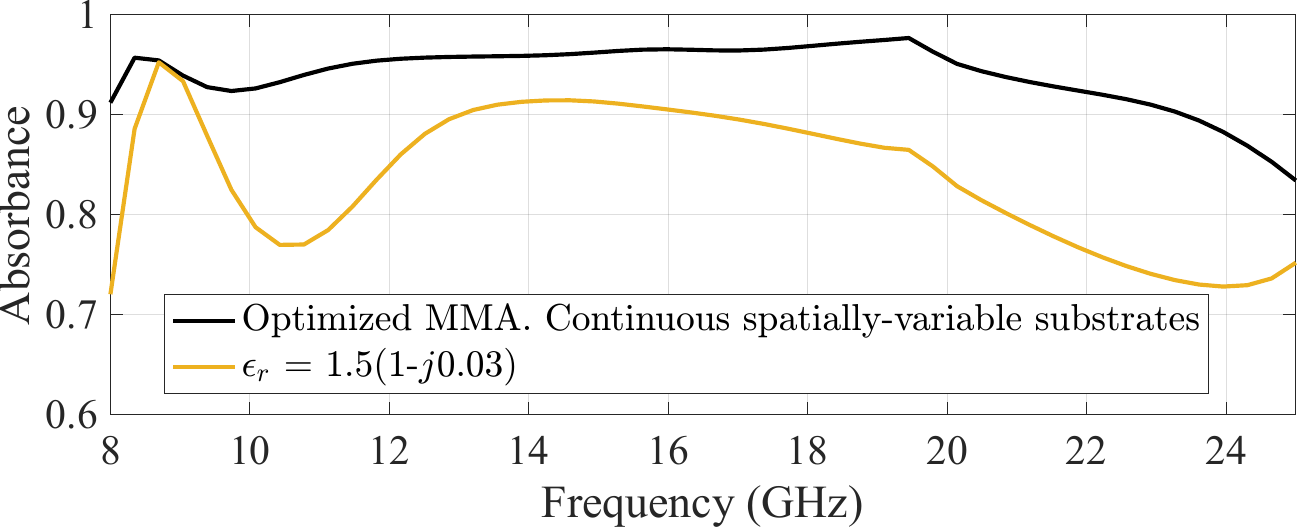}
    \caption{Absorbance of the MXene MMA with (i) constant and (ii) optimized continuous spatially variable dielectric substrates.}
    \label{fig:Absorbance_const_opt_cont}
\end{figure}

\section {MXene-based MMA with spatially variable dielectric substrates}

We demonstrate the \texttt{LOCABINNACONN3D} methodology by synthesizing a multilayer MXene-based MMA with spatially variable dielectric substrates. We start with an MMA of constant dielectric substrates, shown in Fig.~\ref{fig:geom_constant}.  The thickness of each dielectric layer is $h$~=~0.6~mm and the substrate's dielectric constant is $\epsilon_r$~=~1.5(1-$j$0.03). The MMA is metal backed; the ground is modeled as a perfect electric conductor (PEC). The dimensions of the center patch are $W_x = N~p_x$ and $W_y = N~p_y$ along the $x$ and $y$ axes, where $N$ is an integer between 10 and 20, $p_x=D_x/22=0.4618$~mm, $p_y=D_y/48=0.4763$~mm are the cell sizes along the $x$ and $y$ axes, and $D_x$~=~10.16~mm, $D_y$~=~22.86~mm.  The off-center patches have the same length $W_x$ as the center patch along the $x$ axis and half the width $W_y$ along the $y$ axis. 
MXene samples of thickness 0.9$\rm \mu$m, deposited on a polycarbonate membrane of thickness $20\mu$m, were fabricated by vacuum-assisted filtration of Ti$_3$C$_2$T$_x$ aqueous dispersions.  Their surface resistance was characterized in the X-band, using a vector network analyzer (VNA) and WR90 waveguide components; hence, $R_s$~=~12~$\Omega$/sq.  We consider PEC side boundaries and a TE$_{10}$ mode excitation, to resemble a waveguide-fed excitation. We use topology optimization, by the TopOpt method~\cite{Christiansen2021}, to design the MMA's spatially variable substrates, with each cell of size $p_x \times p_y$ having material properties in the continuous range between air and the 3D printer’s resin ($\epsilon_r$~=~2.7). We aim to maximize absorbance in the range $f$=[8,25]~GHz.  The MMA with continuous spatially variable dielectric substrates is shown in Fig.~\ref{fig:opt}. The absorbance of the optimized spatially variable MMA is compared to that of a constant dielectric profile in Fig.~\ref{fig:Absorbance_const_opt_cont}. We observe that the spatially variable substrates enhance the absorption bandwidth. We will transform the optimized MMA into a manufacturable one by the computationally tractable \texttt{LOCABINACONN3D} methodology.

\section{The \texttt{LOCABINACONN3D} methodology}

The \texttt{LOCABINNACONN3D} takes an optimized device with spatially variable substrates, whose cells have material properties in the continuous range between air and the 3D printer’s resin, and transforms it into a manufacturable one, that consists of air and resin, while being computationally tractable. Preliminary results of a simpler \texttt{LOCABINACONN} have been presented in \cite{Passia2025APS}, however they were applied to 2D gratings without either incorporating multilayer fabrication constraints or using semi-analytical solvers. 
We will present the main parts of \texttt{LOCABINNACONN3D} in this section and apply it to the MMA synthesis in the following section. 

\subsection{Discrete materials}

\begin{figure}
    \centering
    \includegraphics[width=1\linewidth]{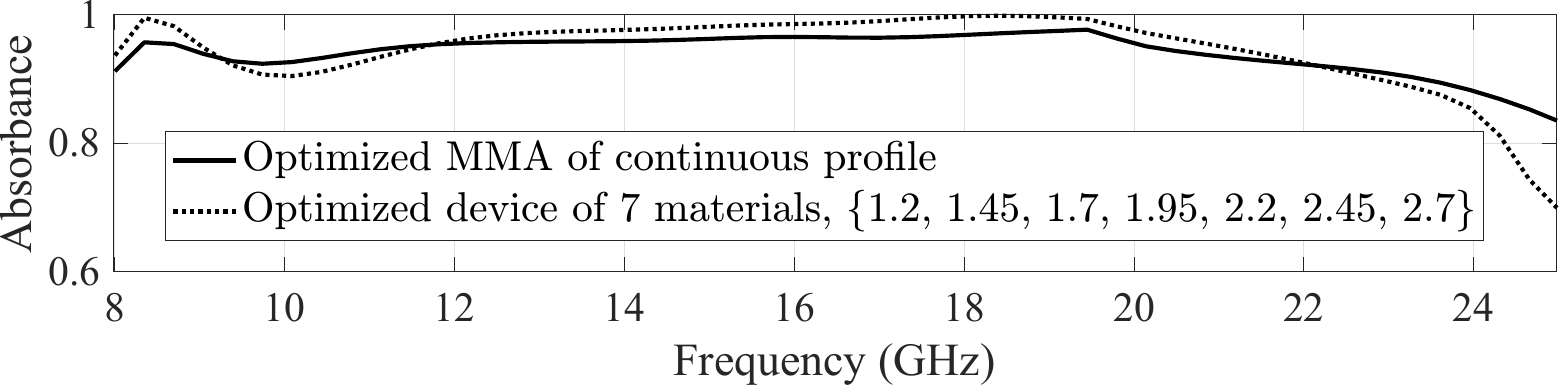}
    \caption{Absorbance of the optimized spatially variable MXene MMA with (i) 7 materials and (ii) continuous dielectric substrates.}
    \label{fig:Abs_7_materials}
\end{figure}

A necessary pre-processing step to implement \texttt{LOCABINNACONN3D} is to transform the  optimized MMA with continuous spatially variable substrates into one with a discrete number of dielectric constant values. Each cell's dielectric constant value is substituted by the closest discrete value. We choose to preserve 7 material levels, as this is sufficient for retaining the absorbance of the optimized continuously-varying spatially variable MMA (Fig.~\ref{fig:Abs_7_materials}).

\subsection{Assigning a resin percentage}

We assign a resin percentage to each material level instead of an air/resin unit cell of predetermined shape. This provides the  design flexibility needed to preserve the absorbance of the continuous spatially varying MMA.  For each material level, we assign the resin percentage for which the propagation constant of the mode supported by the air/resin configuration matches that of the intermediate material level. The \texttt{LOCABINNACONN3D} assigns air or resin to each MMA cell, according to the prescribed air/resin percentage.

\subsection{Manufacturing constraints}

To facilitate multilayer MMA fabrication and assembly, we identify suitable sub-devices that must remain connected. Synthesizing the multilayer MMA as a single connected device is not an option, as the MXene resonators must be inserted between consecutive dielectric layers. To that end, a sub-device is synthesized by selectively extruding regions of each lower dielectric layer and forming complementary gaps in the adjacent upper layer to ensure precise vertical alignment and structural integration. The extruded regions are defined to accommodate the insertion and precise positioning of the MXene patches at designated locations between adjacent dielectric layers. 

\begin{figure}
    \centering
    \includegraphics[width=0.7\linewidth]{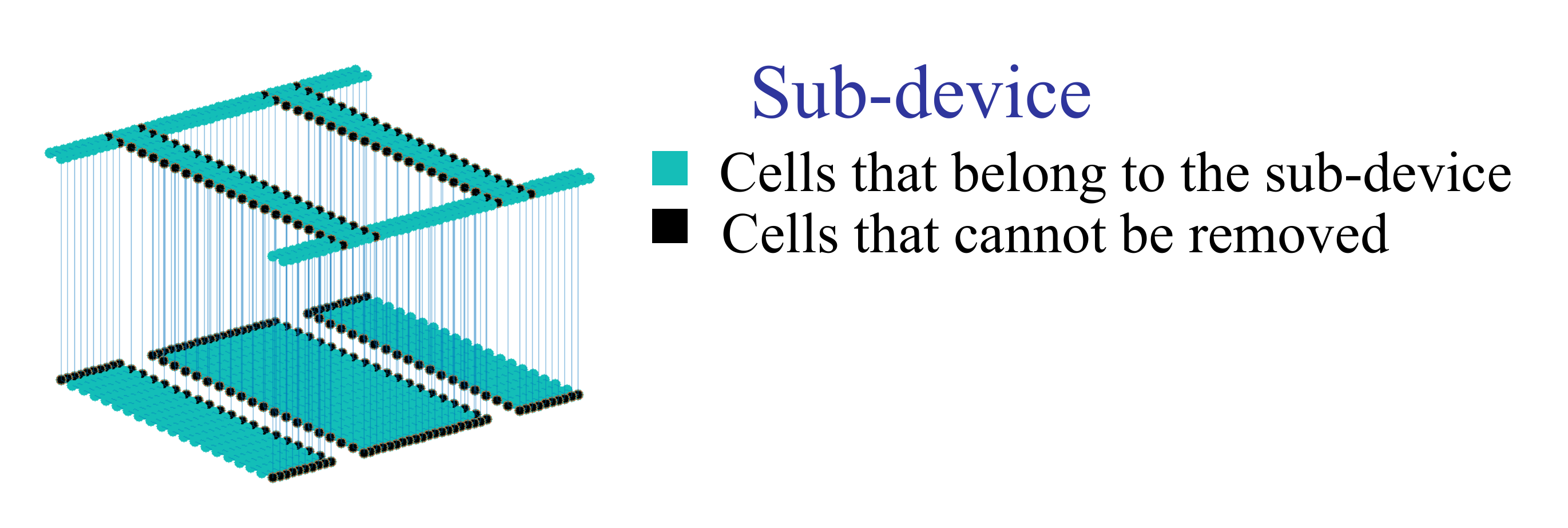}
    \caption{Graph of a sub-device. Each cell of the sub-device is shown as a node (turquoise). Nodes that are immediate neighbors are connected by an edge. The black nodes will remain as resin.}
    \label{fig:Subcomponents_1_2}
\end{figure}

We identify the cells that belong to each sub-device, by treating each cell as a graph node $V$. Two graph nodes are connected by an edge $E$ if they are immediate neighbors (north, west, south, east). 
For the 7-materials non-manufacturable MMA, we form a graph $G(V,E)$ of all cells that belong to a sub-device. 
The graph of a sub-device  is shown in Fig.~\ref{fig:Subcomponents_1_2}. Each cell that belongs to the sub-device is shown as a node (turquoise). The nodes that are immediate neighbors are connected by an edge in Fig.~\ref{fig:Subcomponents_1_2}. 
Our goal is to transform the non-manufacturable sub-devices into manufacturable ones, preserving the connectivity of each sub-device. To promote structural integrity, we specify certain cells that must remain as resin, i.e., the black nodes in Fig.~\ref{fig:Subcomponents_1_2}. This will facilitate the assembly process of consecutive dielectric layers.

\subsection{Computationally-tractable selection of air/resin structures}
\begin{figure}
  \centering
    \subfloat[]{\label{fig:graph_comp_non_manuf}%
    \includegraphics[width = 0.44\columnwidth]{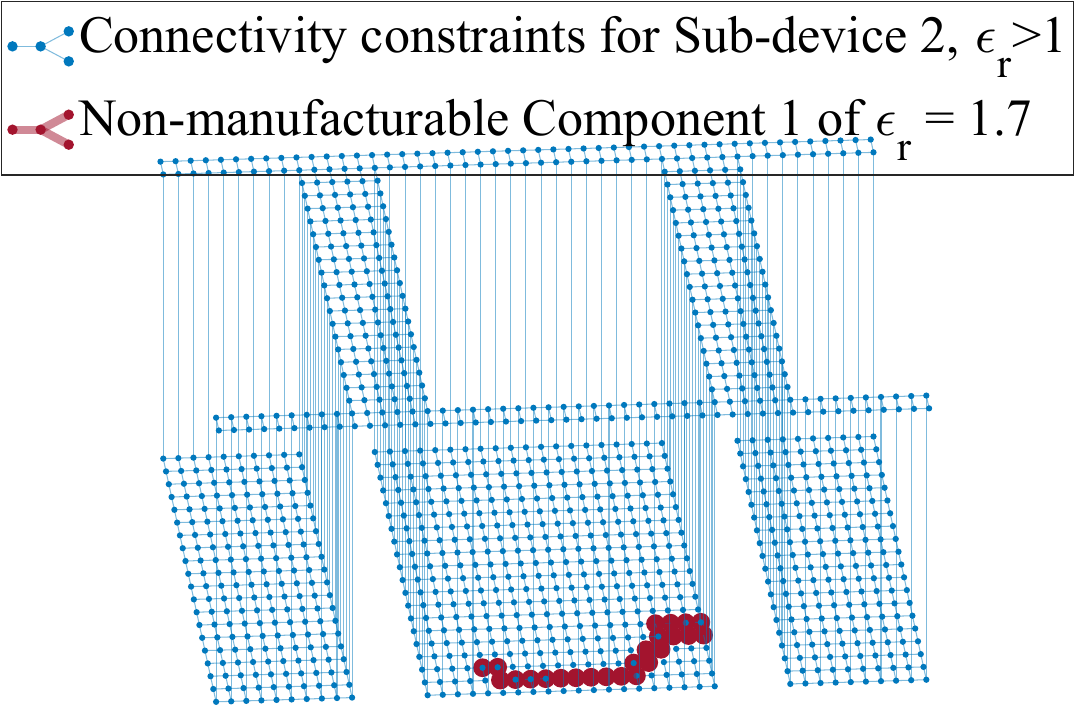}
    }
    \subfloat[]{\label{fig:graph_comp_manuf}%
    \includegraphics[width = 0.44\columnwidth]{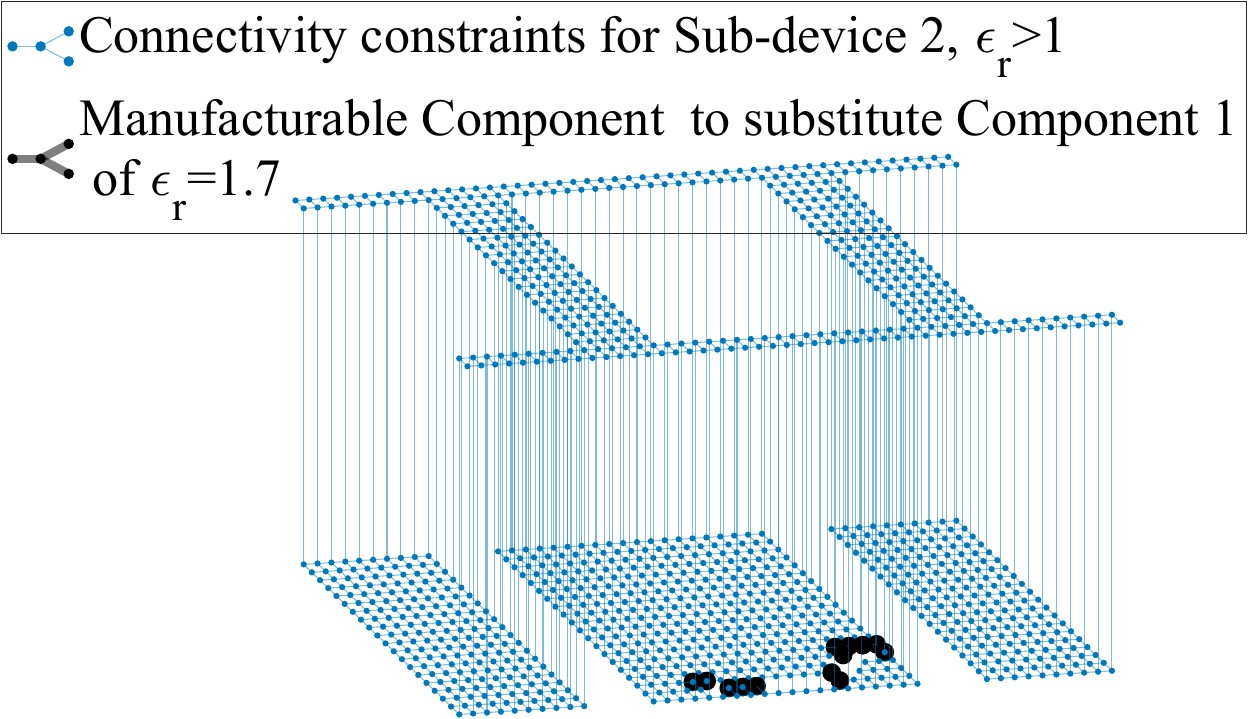}
      }   
    \caption{Graph of the non-air nodes and edges of the second sub-device (blue color), that will be printed as a single unified structure.  (a) The red nodes and edge correspond to the cells of Component 1 of material $\epsilon_{r,d}$=1.7. (b) The nodes of (a) are substituted by an air/resin structure. The black correspond to resin, after the \texttt{LOCABINACONN3D} is applied.}
\end{figure}

\begin{figure}
  \centering
    \subfloat[]{\label{fig:S11_comp}%
    \includegraphics[width = 1\columnwidth]{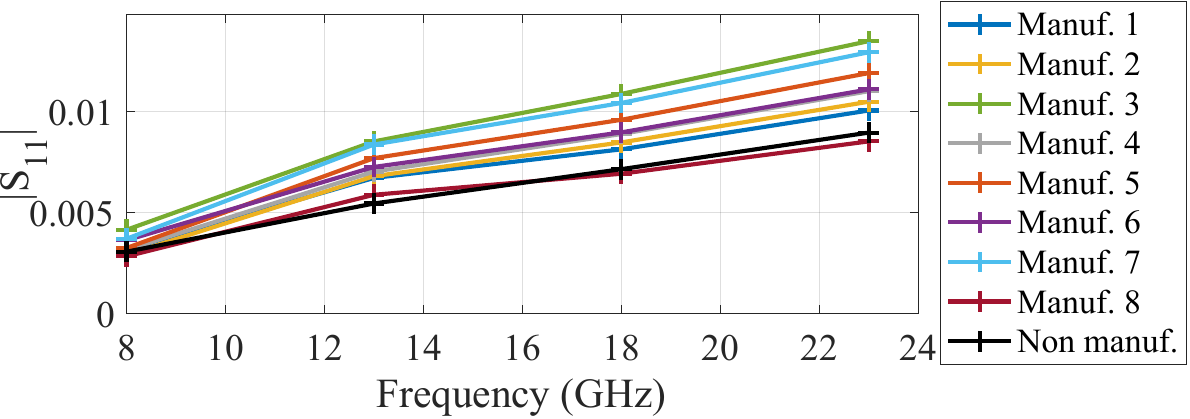}
    }
    \\
    \subfloat[]{\label{fig:S21_comp}%
    \includegraphics[width = 1\columnwidth]{ 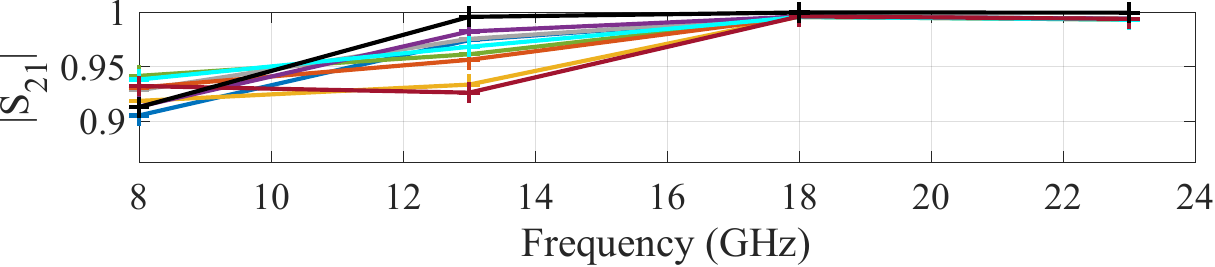}
      }   
    \caption{(a) $|S_{11}|$ and (b) $|S_{21}|$ of the non-manufacturable structure of 7 materials of ``Sub-device 2 - $\epsilon_r$=1.2 - Component=1'' and 8 manufacturable structures.}
\end{figure}

We implement the \texttt{LOCABINNACONN3D} algorithm for each sub-device, with connectivity constraints imposed on a
sub-device level.
The steps of the methodology are as follows:
\begin{list}{$\bullet$}{%
      \setlength{\leftmargin}{0em}
      \setlength{\labelwidth}{1.5em}
      \setlength{\labelsep}{0.5em}
      \setlength{\itemsep}{0.0ex}
      \setlength{\parsep}{0pt}
      \setlength{\itemindent}{\dimexpr\labelwidth+\labelsep\relax}%
      \setlength{\topsep}{0ex}%
  }%
\item For each material level $\epsilon_{\rm r,d}$, we form the corresponding graph $G_d$ and locate its graph components, i.e., material regions that are intrinsically connected. As an example, we show in red  in Fig.~\ref{fig:graph_comp_non_manuf} the nodes and edges of non-manufacturable component 1 of the material $\epsilon_{\rm r,d}$=1.7 in the second sub-device.
\item  For each graph component $i$ of the material $\epsilon_{\rm r,d}$, we generate multiple manufacturable air/resin configurations. A manufacturable air/resin structure is shown in Fig.~\ref{fig:graph_comp_manuf}, which serves as a candidate for substituting the non-manufacturable component of Fig.~\ref{fig:graph_comp_non_manuf}. The nodes in black correspond to the remaining resin cells whereas white areas are air.
  \begin{list}{$-$}{%
              \setlength{\leftmargin}{1.0em}
              \setlength{\labelwidth}{0.0em}
              \setlength{\labelsep}{0.3em}
              \setlength{\itemsep}{0.0ex}
              \setlength{\parsep}{0pt}
              \setlength{\itemindent}{\dimexpr\labelwidth+\labelsep\relax}%
              \setlength{\topsep}{0ex}%
          }%
  \item To generate each manufacturable air/resin configuration, we set $n_{a}$ cells to air, according to the air/resin percentage.  
  \item A uniform random distribution is used to determine the cells that will be set to air in each graph component. We want air and resin to be distributed more uniformly, i.e., to have configurations with finer features.   
  \item Before setting a cell to air, we check that the MMA sub-device remains connected upon removal of the cell. If it disconnects, we return to the previous configuration.  We also check that the cell does not belong to those that must remain as resin (black nodes of Fig.~\ref{fig:Subcomponents_1_2}).
  \end{list}   
\item For each graph component $i$, we select the manufacturable configuration that achieves the closest response to the existing non-manufacturable configuration.
                \begin{list}{$-$}{%
              \setlength{\leftmargin}{1.0em}
              \setlength{\labelwidth}{0.0em}
              \setlength{\labelsep}{0.3em}
              \setlength{\itemsep}{0.0ex}
              \setlength{\parsep}{0pt}
              \setlength{\itemindent}{\dimexpr\labelwidth+\labelsep\relax}%
              \setlength{\topsep}{0ex}%
          }%
      \item The selection is made by simulating the non-manufacturable and all different manufacturable structures that represent graph component $i$. 
      \item We calculate the S-parameters of each structure and select the manufacturable structure that provides the closest match to the non-manufacturable one.
      \item Instead of a full-wave solver, such as the FEM, we use a method-of-lines (MoL) solver~\cite{Bideskan2023, Alu2004,Passia2025}, which provides a faster alternative. The MoL is a semi-analytical method, where the wave equations are discretized only on the transverse $x-y$ plane by a 2-D discretization scheme based on the Yee cell, and are solved analytically along the third dimension, which considerably decreases the degrees of freedom (DoFs). Using the MoL, we calculate the transfer matrix of each structure's layer and multiply the transfer matrices of consecutive layers to obtain the transfer matrix of the entire manufacturable and non-manufacturable dielectric structures. Finally, we calculate the scattering matrix of the entire dielectric structure and  obtain its S-parameters. To implement the MoL, we form a box of size $N_x \times N_y \times N_z$ that encloses the dielectric structure. The box cells that do not belong to the component  are set to air. We impose PEC boundary conditions to the side boundaries. The $|S_{11}|$ and $|S_{21}|$ of the non-manufacturable and eight manufacturable structures are shown in Fig.~\ref{fig:S11_comp} and Fig.~\ref{fig:S21_comp} for ``Sub-device 2, material $\epsilon_r=1.2$,  component 1''. The computational box, used for this component in the MoL, is of size $22 \times 13 \times 2$, which is considerably smaller than the entire MMA of size $22 \times 48 \times 7$.  We select manufacturable structure 1 as the best match for both $|S_{11}|$ and $|S_{21}|$. 
      \end{list}
\item We combine all these smaller manufacturable structures to form the entire manufacturable MMA.
\end{list}


\section{Assessment of the manufacturable MMA}
\begin{figure}
  \centering
    \includegraphics[width = 1\columnwidth]{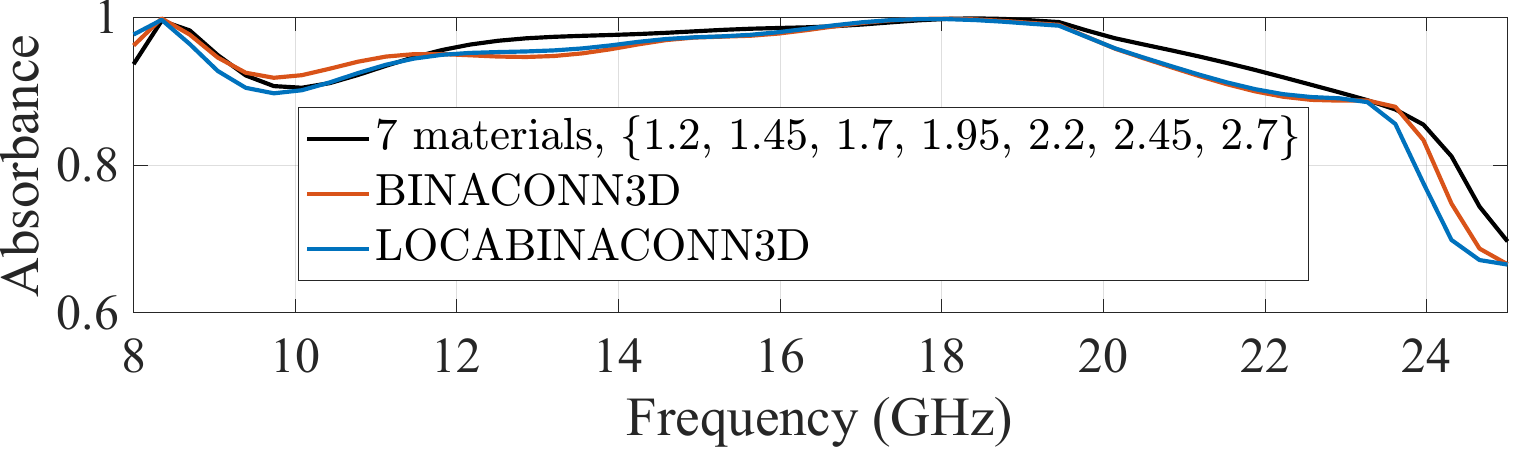}
    \caption{Absorbance of the manufacturable MMA  by LOCABINACONN3D \& BINACONN3D, and non-manufacturable MMA.}
    \label{fig:A_manufacturable_non_manufacturable}%
\end{figure}

We apply the \texttt{LOCABINNACONN3D} method on all sub-devices, materials, and components, and combine all manufacturable structures to form the manufacturable MMA. We simulate the manufacturable MMA obtained by \texttt{LOCABINNACONN3D} in COMSOL Multiphysics and compare the absorbance to that of the non-manufacturable MMA of seven materials and the manufacturable MMA obtained by \texttt{BINNACONN3D}. The absorbance of all three cases is shown in Fig.~\ref{fig:A_manufacturable_non_manufacturable}. The \texttt{LOCABINNACONN3D}-based manufacturable MMA achieves a satisfactorily close absorbance to both the \texttt{BINNACONN3D}-based manufacturable MMA and the non-manufacturable 7-material level MMA. Specifically, the average absorbance in the range 8-25~GHz is $A$=0.9301 for \texttt{LOCABINNACONN3D}-based manufacturable MMA, $A$=0.9344 for \texttt{BINNACONN3D}-based manufacturable MMA, and $A$=0.9439 for the non-manufacturable 7-material MMA. An FEM simulation of the entire manufacturable MMA, used in \texttt{BINNACONN3D}, has about 630,000 DoFs and takes about 44 seconds per frequency per device on an AMD Ryzen Threadripper 3960X system (24 cores, 48 threads, 3.80 GHz, DDR4-3200).  The MoL simulations of the smaller components take 0.36 seconds on average per frequency per configuration.  The \texttt{LOCABINNACONN3D} can be applied to even larger-scale MMAs where simulating the entire device would be even more computationally demanding.

\section{Conclusions}
We introduced the \texttt{LOCABINNACONN3D} methodology, which enables the computationally tractable and scalable synthesis of optimized MXene-based MMAs with spatially variable dielectric substrates. The manufacturable optimized MMA maintains the absobance of the non-manufacturable one. By simulating considerably smaller manufacturable MMAs and using MoL instead of FEM, our work paves the way for the synthesis of larger-scale optimized metamaterial devices.

\section*{Acknowledgment}
This project has received funding from the European Union's Horizon 2020
research and innovation programme under the Marie Skłodowska-Curie grant
agreement No.101146306.
\protect\usebox{\fundinglogo}

\bibliographystyle{IEEEtran}

\bibliography{ref} 

\end{document}